\documentclass[prb,twocolumn,showpacs,preprintnumbers,superscriptaddress,amsmath,amssymb]{revtex4}
\usepackage{dcolumn}
\usepackage{epsfig}
\usepackage{times}

\usepackage{color}

\definecolor{Blue}{rgb}{0,0,1}
\definecolor{NavyBlue}{rgb}{0.14,0.14,0.56}
\definecolor{rot}{cmyk}{0,1,1,0}

\begin{document}
\def\bra#1{\mathinner{\langle{#1}|}}
\def\ket#1{\mathinner{|{#1}\rangle}}
\def\braket#1{\mathinner{\langle{#1}\rangle}}
\def\Bra#1{\left<#1\right|}
\def\Ket#1{\left|#1\right>}
\def\bravert{\egroup\,\vrule\,\bgroup}
\newcommand{\e}[1]{\cdot 10^{#1}}
\newcommand{\wn}{\,cm$^{-1}$}
\newcommand{\ea}{\emph{et al.}}
\newcommand{\dg}{$^{\circ}$}

\title{The  phonon dispersion of graphite by inelastic x-ray scattering}
\author{M. Mohr}
 \email{marcel@physik.tu-berlin.de}
\affiliation{Institut f\"ur Festk\"orperphysik, Technische Universit\"at Berlin,
Hardenbergstr. 36, 10623 Berlin, Germany}

\author{J. Maultzsch}
\thanks{present address: Departments of Electrical Engineering and Physics, Columbia University, New York NY 10027, USA}
\affiliation{Institut f\"ur Festk\"orperphysik, Technische Universit\"at Berlin,
Hardenbergstr. 36, 10623 Berlin, Germany}

\author{E. Dobard\v zi\'c}
\affiliation{Faculty of Physics, University of Belgrade, POB 368, 11011
  Belgrade, Serbia}

\author{S. Reich}
\affiliation{Department of Materials Science and
Engineering, Massachusetts Institute of Technology, 77 Massachusetts Avenue,
Cambridge, MA 02139-4307, USA}

\author{I. Milo\v sevi\'c}
\author{M. Damnjanovi\'c}
\affiliation{Faculty of Physics, University of Belgrade, POB 368, 11011
  Belgrade, Serbia}

\author{ A. Bosak }
\affiliation{European Synchrotron Radiation Facility (ESRF), BP 220, F-38043 Grenoble cedex, France}
\author{M. Krisch  }
\affiliation{European Synchrotron Radiation Facility (ESRF), BP 220, F-38043 Grenoble cedex, France}

\author{C. Thomsen}%
\affiliation{Institut f\"ur Festk\"orperphysik, Technische Universit\"at Berlin,
Hardenbergstr. 36, 10623 Berlin, Germany}

\date{\today}

\begin{abstract}

We present the full in-plane phonon dispersion of graphite obtained from inelastic x-ray scattering, including the optical and acoustic branches, as well as 
  the mid-frequency range between the
$K$ and $M$ points in the Brillouin zone, where experimental data have been unavailable so far. The existence of a Kohn anomaly at the $K$ point is further
 supported. We fit a fifth-nearest neighbour force-constants model 
to the experimental data, making improved force-constants calculations of the phonon dispersion in both graphite and carbon nanotubes available.

\end{abstract}


\maketitle

\section{Introduction}

Research on carbon nanotubes and the recent availability of single graphite sheets\cite{novoselov04,zhang05} (graphene) has revived the  interest in the fundamental physical properties of graphite during the last years. Carbon nanotubes can be  regarded as one or more rolled-up
graphite sheets, and many physical
properties of carbon nanotubes are closely related to those of graphite.\cite{reich04buch,thomsen07}

The fundamental characteristics of a crystalline material comprise its phonon spectrum, from which one can derive several other physical properties such as sound velocity, thermal
conductivity, or heat capacity. Furthermore, phonons play an important role in excited-state dynamics and electrical transport properties. Optical or electronic excitations can decay into vibrational excitations or can be scattered by phonons into different states. For example,
in carbon nanotubes the high-bias electrical transport is assumed to be limited by scattering of the carriers by
optical phonons corresponding to the graphite $K$ point.\cite{park04,lazzeri06} 

\begin{figure}[t]
\centering
 \epsfig{file=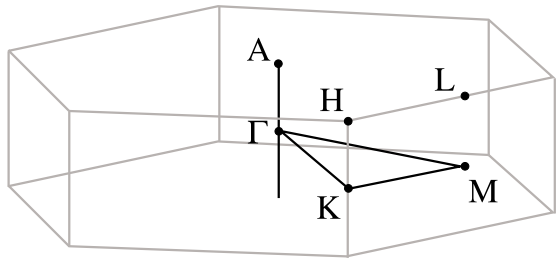,width=7.3cm}
\caption{\label{bild:bri_zone} Brillouin zone of graphite. The Brillouin zone of graphene is the hexagon which lies in the plane with the points
 $\Gamma ,K$ and $M$. The distances between the high-symmetry points are $\Gamma-K=4\pi/3a$, $\Gamma-M=2\pi/\sqrt{3}a$, and $K-M=2\pi/3a$.
}
\end{figure}

The phonon dispersion of graphite has not been
completely resolved by experiment, mostly due to the lack of large enough samples of crystalline quality.
It has been partly  measured by inelastic neutron scattering (INS), electron-energy loss spectroscopy (EELS),
and  inelastic  x-ray scattering (IXS).\cite{nicklow72,oshima88,siebentritt97,yanagisawa05,maultzsch04}
Most experiments so far have determined the dispersion along the $\Gamma-K$ and the $\Gamma-M$ directions in the graphite Brillouin zone (see Fig.~\ref{bild:bri_zone} for a definition of the Brillouin zone).
The recent measurement of the optical branches  along the $K-M$ direction by IXS pointed to the existence of a Kohn anomaly for the highest phonon branch at the $K$ point.\cite{maultzsch04,piscanec04}
Although this result resolved previous discrepancies between different lattice dynamics models,
there are still open questions  regarding the
shape of lower-lying phonon branches.
In particular, differences
appear between force-constants  and density-functional theory (DFT) calculations, where experimental data are still unavailable. This concerns, e.g., the crossing between the acoustic and optical bands near the $M$-point or the energy of the
transverse acoustic mode at the $K$ point.
For carbon nanotubes, 
the experimental determination of the  phonon dispersion throughout the entire Brillouin zone would require
monocrystalline samples  of a minimum size, which have been unavailable so far. Therefore, 
the closest approximation to the experimental phonon dispersion of carbon 
nanotubes is currently the phonon dispersion of graphite.

Here we present  the  phonon dispersion of graphite in all three high-symmetry directions in the basal plane
determined by inelastic x-ray scattering. In particular, the phonon branches between the $K$ and $M$ point and the acoustic branches in all high-symmetry directions are obtained, giving
both the optical and acoustic phonons from one experimental technique. We fit our data by a set of force constants, including   fifth-nearest neighbours of carbon atoms. The fitted force constants can be used to deduce the corresponding force constants for carbon nanotubes. 

This  paper is organized as follows.
In the next section we briefly describe the experimental details of the IXS experiments. We give an introduction to
the phonon dispersion of graphite and present the experimental data in Sect.~\ref{sec:results}.
In Sect.~\ref{sec:force-constant-fits} we apply a fifth-nearest neighbour force constants fit to the experimental data and provide the in-plane and out-of-plane force constants.  

\section{Experimental setup}
\label{sec:experiment}

The inelastic x-ray  
experiments were performed at  beam line ID28 at the European
Synchrotron Radiation Facility (ESRF). For a review of IXS the reader is
referred to Refs.~\onlinecite{burkel00} and \onlinecite{krisch07}. The energy of the incident radiation of 17794\,eV was selected by the  (999) Bragg reflection of a silicon crystal.
The scattered photons were analyzed by five analyzers operating in the same reflection
order. The total energy resolution in this configuration is 3.0\,meV.\cite{krisch07}
The x-ray beam was focused to $250\times 60\,\mu$m$^2$, selecting a single microcrystal in a
naturally grown graphite flake. The typical size of a single grain was about 800 $\mu$m in lateral direction and 100$\mu$m
along the $c$-axis. By x-ray diffraction
we obtained the lattice parameters $a=2.463$\,\AA \ and $c=6.712$\,\AA , in excellent agreement with previous
neutron diffraction data ($a=2.464$\,\AA, $c=6.711$\,\AA).\cite{trucano75}

Inelastic scattering spectra were recorded by varying the temperature difference between the monochromator and the analyzer silicon crystal.
To minimize the effects of temperature drifts that could result in an energy offset, we performed systematic
Stokes--anti-Stokes scans between the measurements.
In our setup  the $c$-axis of graphite and the scattering
plane encompasses an angle of 90\dg , 30\dg , and 0\dg, depending on the phonon branch under
consideration.
The scattering geometry was chosen according to the selection rules, see Ref.~\onlinecite{kirov03}.

\section{Experimental results}
\label{sec:results}

The unit cell of graphene contains two atoms, resulting in six phonon
branches. The unit cell of graphite consists of four atoms, which leads to twelve phonon branches. The space group of graphite is
$P6_3/mmc$ (international notation). At the $\Gamma$ point it possesses the factor group 6/mmm 
($D_{6h}$ in Sch\"onfliess notation).
The optical zone-center modes of graphene are decomposed into $\Gamma=B_{2g}+E_{2g}$.
In graphite, the optical zone center modes are decomposed into  $\Gamma=A_{2u}+2B_{2g}+E_{1u}+2E_{2g}$.\cite{nemanich79,inuibuch,reich04}
The $A_{2u}$ and $E_{1u}$ modes are IR active, the $E_{2g}$ modes Raman active. The $B_{2g}$ modes are optically inactive, but
can be measured via INS or IXS.
The three acoustic modes are decomposed into $\Gamma=A_{2u}+E_{1u}$.

Graphite is a highly anisotropic material: the nearest-neighbor distance between two atoms in the plane is
$a/\sqrt{3}\approx1.42$\,\AA , while the inter-layer distance is $c/2\approx3.35$\,\AA .
The bonds between two carbon atoms in the plane are much stronger than the weak van-der-Waals interactions between the layers. Therefore, 
compared to graphene, one expects that the phonon modes of graphite
correspond approximately to in-phase and out-of-phase vibrations of the two graphene planes. 
Most of the phonon 
branches in  graphite are nearly doubly degenerate and almost the same as in graphene.\cite{pavone96,piscanec04} Only
close to the $\Gamma$ point, the acoustic modes of the single layer split in graphite into an acoustic mode (in-phase vibration of the
graphene sheets) and an optical mode [out-of-phase vibration; in-plane: $E_{2g}$ at 5.2\,meV (42\,cm$^{-1}$); out-of-plane:
$B_{2g}$ at 15.7\,meV (127\,cm$^{-1}$)]. 
For the optical modes of graphite, the difference between the in-phase and the out-of-phase vibrations is very small: at the $\Gamma$ point the IR active $E_{1u}$ mode  is found  at 196.9\,meV (1588\,\wn), close to the Raman active $E_{2g}$ mode at 196.0\,meV (1581\,\wn).
The same holds for  the $A_{2u}$ mode at 107.5\,meV (867\,\wn)\cite{nemanich77b,nemanich79} and the $B_{2g}$ mode at 107.6\,meV (868\,\wn).
Therefore, in the following theoretical discussion, we will consider the phonons of a single graphene sheet.

The six branches are divided into the out-of-plane acoustic mode ZA, the in-plane acoustic mode TA (sometimes called SH=shear), the longitudinal acoustic mode
LA, the out-of-plane optical mode ZO, the in-plane optical mode TO (SH$^*$), and the longitudinal optical mode LO.
Four branches belong to modes where the atoms move in-plane with the graphene layer (TA, LA, TO, LO); two branches belong to
transverse modes, where the atoms move  out of the plane (ZA, ZO).

\begin{figure*}[t]
\centering
 \epsfig{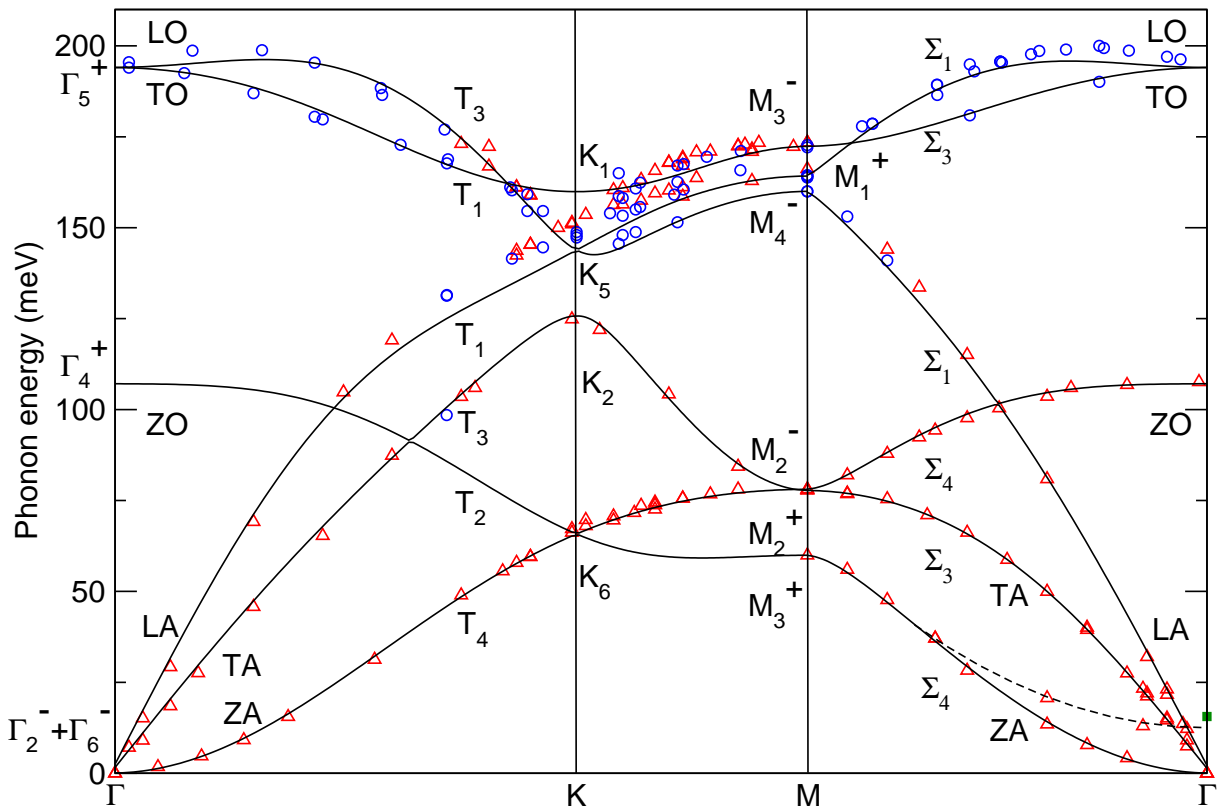}
\caption{\label{bild:bands} (Color online)
Phonon dispersion of graphite from inelastic x-ray scattering (symbols). 
Triangles are present data, circles are data already published in Ref.~\onlinecite{maultzsch04}.
The full square at the $\Gamma$ point is INS data from Ref.\onlinecite{nicklow72}. Solid lines are the
force-constants calculations from the 5th-nearest neighbor fit discussed in Sect.~\ref{sec:force-constant-fits}; the
dashed line is a quadratic extrapolation of the data. The lines are denoted by their symmetry representation in
space group notation. The relation between space group and molecular notation can be found in Table~\ref{tab:spacegrp}.
}
\end{figure*}

In Fig. \ref{bild:bands} we show our experimental data of the graphite phonon dispersion in the plane.
The lines show the fifth-nearest neighbor force-constants fit described in Sect.~\ref{sec:force-constant-fits}.
The optical phonon frequencies near the $\Gamma$ point agree well with
previous experiments. We find the $E_{2g}$ LO mode  at 196.0\,meV (1581\,cm$^{-1}$), and the $B_{2g}$
mode at 107.6\,meV (868\,cm$^{-1}$). 
Regarding the overall shape of the phonon branches, our experiments confirm previous \emph{ab-initio} DFT calculations, letting aside the special situation for the highest branch at the $K$ point.\cite{pavone96,sanchez99a,dubay03,maultzsch04,piscanec04}

 As can be seen, the highest optical frequency does not appear at the $\Gamma$-point. Instead, the phonon frequency first increases with larger wave vector and then decreases again. This effect, called overbending, has been observed in diamond as well.\cite{kulda02}
In graphite, it has been predicted to result from a Kohn
anomaly, i.e., the frequency at the $\Gamma$ point is lowered due to interaction of the phonon with the electronic
system.\cite{maultzsch04,piscanec04} Another Kohn anomaly in graphite can be found for the TO-derived phonon branch at the $K$-point (fully symmetric $A_1^\prime$ ($K_1$) mode).
We have gained additional  data for the highest optical phonons around the $K$ point, confirming previous measurements of the frequency softening of the TO-derived branch near the $K$ point.\cite{maultzsch04} Again, we were not able to detect the  $A_1^\prime$ phonon directly at the $K$ point.
The strong electron-phonon interaction has been predicted to reduce the phonon lifetime which results in a line broadening. Probably the
large line width makes it very difficult to detect the $A_1^\prime$  phonon at the $K$ point experimentally. 

\begin{figure}[t]
\centering
\epsfig{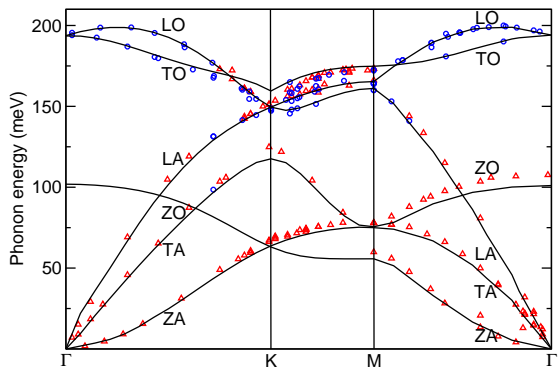}
\caption{\label{bild:bands_DFT} (Color online)  Our in-plane phonon dispersion of graphite together with a DFT calculation from
  Ref.\onlinecite{maultzsch04} (solid lines). Same symbols as in Fig.~\ref{bild:bands} were used.
}
\end{figure}

 Considering the  differences between previous theoretical models, we find the following results, see also Sect.~\ref{sec:force-constant-fits}. Between the $\Gamma$ and $M$ points, the ZO and TA modes  do not cross within our experimental error of
3\,meV. This is in contrast to previous empirical force-constants models and EELS data of Ref.~\onlinecite{oshima88}. Which branch is  higher directly at the  $M$ point cannot be uniquely distinguished from our data.
However, in DFT phonon calculations the crossing is found between $K$ and $M$ but close to $M$ (about 1/10 of the
distance between $K$ and $M$, see Ref.~\onlinecite{dubay03}). The overall agreement with DFT calculations supports the
crossing between $K$ and $M$, see also Fig.~\ref{bild:bands_DFT}. 

The TA branch along $\Gamma -M$ shows a smaller increase compared to the
electron energy loss spectroscopy (EELS) data in Ref.~\onlinecite{oshima88} In a recent
EELS experiment on epitaxially grown thin graphene sheets,\cite{yanagisawa05} however,  this branch could not be detected, as the shear modes in graphene are forbidden in EELS. These contrasting results suggest
that the crystalline quality of Ref.~\onlinecite{oshima88} was lower, softening the selection rules.
This explains why some previous empirical models, relying on the then available EELS data, predicted a larger slope and consequently a crossing of the ZO and TA modes between the $\Gamma$ and $M$ points. 

 We measured for the first time the ZA and TA mode between the $K$ and $M$ points. Our results confirm predictions made by \emph{ab-initio}
calculations very well (Fig.~\ref{bild:bands_DFT}), and are also well reproduced by our force-constant fit (Fig.~\ref{bild:bands}). The trend of both branches crossing
near the $M$ point can be recognized. 

The two optical phonons at the $M$ point derived from the LO and LA
branches are very close in frequency {($<$ 4\,meV )}, and we were not able to distinguish them clearly by symmetry. It appears, however, consistent with DFT calculations and the force constants fit in Sect.~\ref{sec:force-constant-fits} that the higher frequency has $M_{1}^+$ symmetry and the lower one $M_4^-$. As a consequence, the LO- and LA-derived branches cannot cross between the $K$ and $M$ point.

\begin{figure}[t]
\centering
\epsfig{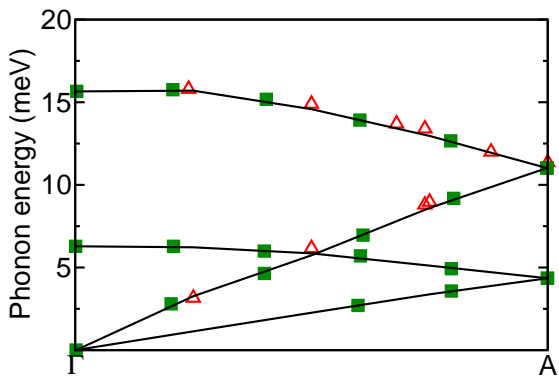}
\caption{\label{bild:bands_GA} (Color online)  Phonon dispersion of graphite along the $\Gamma-A$ direction.
Open triangles are present IXS data, full squares are neutron scattering data from  Ref.~\onlinecite{nicklow72}. The lines are a guide to the eye. 
}
\end{figure}

In Fig.~\ref{bild:bands_GA} we show the low frequency phonon range along the $\Gamma -A$ direction, i.e., perpendicular to the in-plane direction. For comparison, we also present the INS
data on highly oriented pyrolytic graphite from Ref.~\onlinecite{nicklow72}. They are in excellent agreement. The
high-frequency phonon range is expected to show almost no dispersion along the $\Gamma -A$ direction.\cite{pavone96}

Due to experimental reasons we were not able to
record data points from the ZO branch along the $\Gamma-K-M$ direction. This branch has been measured in recent EELS experiments of Ref.~\onlinecite{yanagisawa05}. In general, the   data of Ref.~\onlinecite{yanagisawa05} 
agree well with ours, but at the $K$-point the ZO and ZA branches in Ref.~\onlinecite{yanagisawa05} show a relatively
large splitting of $\approx$10\,meV. They cannot stem from the degenerate $K_6$-phonon, but could possibly represent the
out-of-phase modes of the graphite planes. 
On the other hand, in DFT calculations of graphite\cite{pavone96} this splitting seems much smaller than indicated by the EELS data.
 
Regarding the phonon modes specific for graphite with more than one layer, we find the low-energy out-of-phase modes near the $\Gamma$ point. These are indicated by the dashed line in Fig.~\ref{bild:bands}.
  We measured two out-of-phase ZO$^\prime$ phonons in the $\Gamma-M$ direction, with  energies 13.6\,meV and 23.3\,meV at
  0.16 of the $\Gamma-M$ distance and at 0.4\,$\Gamma-M$, respectively. A quadratic extrapolation leads to a value of 12.5\,meV at the $\Gamma$ point, in
  agreement with 15.7\,meV from neutron scattering data.\cite{nicklow72}

The optical-phonon frequencies at the high-symmetry points  from our experiment  are summarized in
Figs.~\ref{bild:Gamma}, \ref{bild:K}, and \ref{bild:m} together with the displacement patterns obtained by the force-constants calculations in Sect.~\ref{sec:force-constant-fits}. 
The acoustic phonon branches near the $\Gamma$ point give information on the elasticity of graphite, which will be reported elsewhere.\cite{bosak06b}

\section{Force constants calculations}
\label{sec:force-constant-fits}
Phonon dispersion relations are often predicted from  \emph{ab-initio} DFT or from 
empirical force-constants (FC) calculations. Empirical force-constants models in graphite have so far included up to 4th-nearest neighbors, in order to reproduce the overbending of the optical branch near the $\Gamma$ point.\cite{jishi82,jishi93} In the case of graphite, besides the details regarding the frequency values, both methods show
 differences in the shape of phonon branches, e.g., the position of the crossing of the ZO and TA modes near the $M$ point.  While in FC calculations a
crossing of the ZO and TA branches between the $\Gamma$ and $M$ points is predicted, it is found in
\emph{ab-initio} results to take place between $K$ and $M$. This probably stems from a
fit to the only available TA mode from EELS experiments in Ref.~\onlinecite{oshima88}, as our force-constants fit will show later (e.g. see Ref.\onlinecite{jishi93}). 

Further differences between force-constants  and \emph{ab-initio} DFT calculations are found regarding the LA
and LO branches near the $M$ point: In DFT results, the LO-derived phonon branch is higher than the LA  phonon at the $M$-point, and, as a result, the two branches do not cross between $M$ and $K$; \emph{vice versa} in most predictions by empirical force constants.

Discrepancies with the experimental data are found for both models for the TO-derived branch at the $K$  point, except for Ref.~\onlinecite{mapelli99}. 
In this context we want to emphasize the importance of the $K$-point, when performing DFT
calculations.  The atomic forces in graphite are long ranged. Therefore,  when using the finite-differences approach and
DFT~\cite{sanchez99a,dubay03,maultzsch04} only
the phonons commensurate with the supercell are calculated correctly. In linear-response calculations, on the other hand,  the implementation of the $K$-point can be more
easily achieved,\cite{piscanec04,pavone96} however, the dynamical matrix at the $K$-point should be explicitly calculated and not be simply interpolated.

\begin{figure}[t]
\centering
\epsfig{file=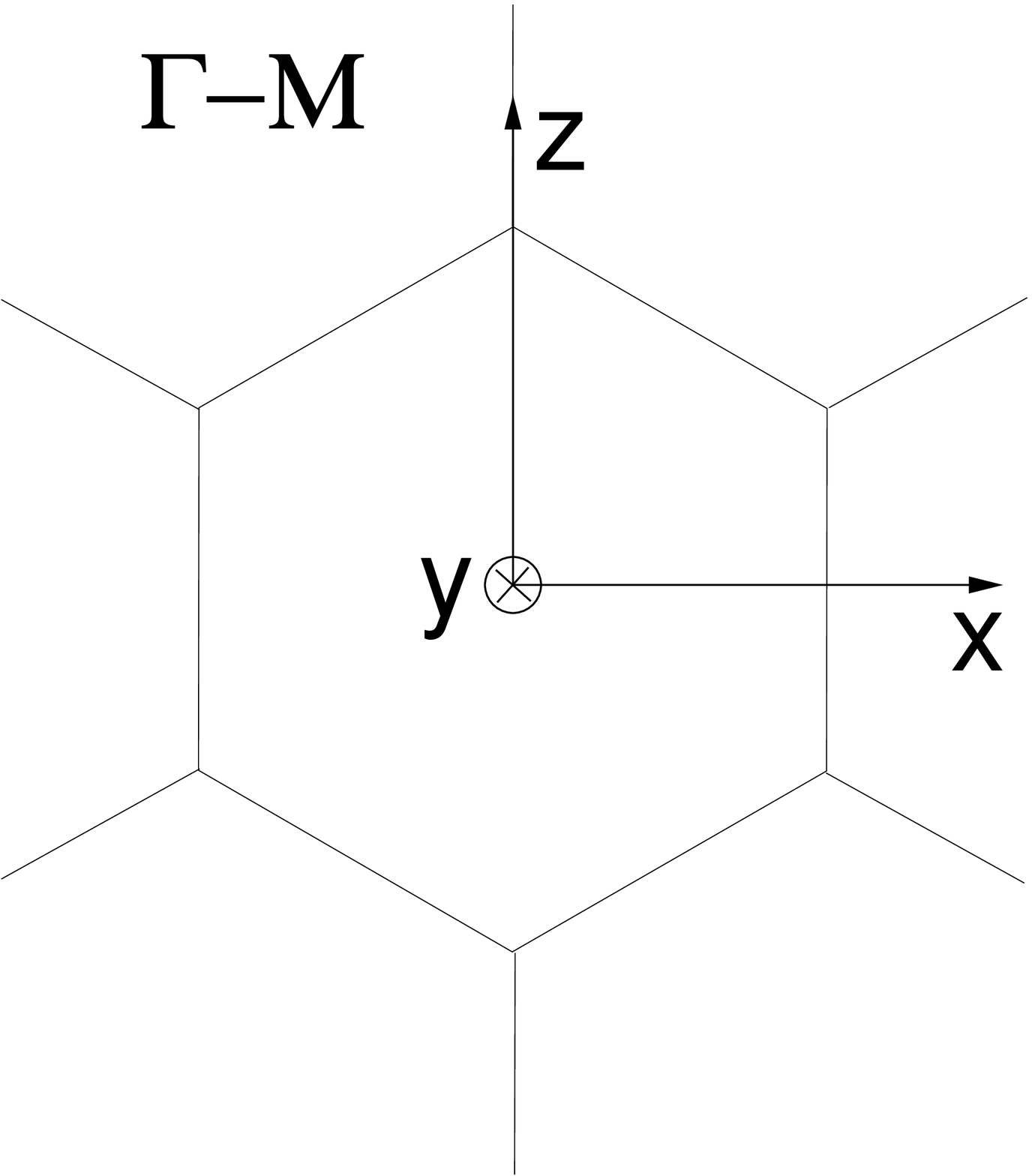,width=2.5cm}
\epsfig{file=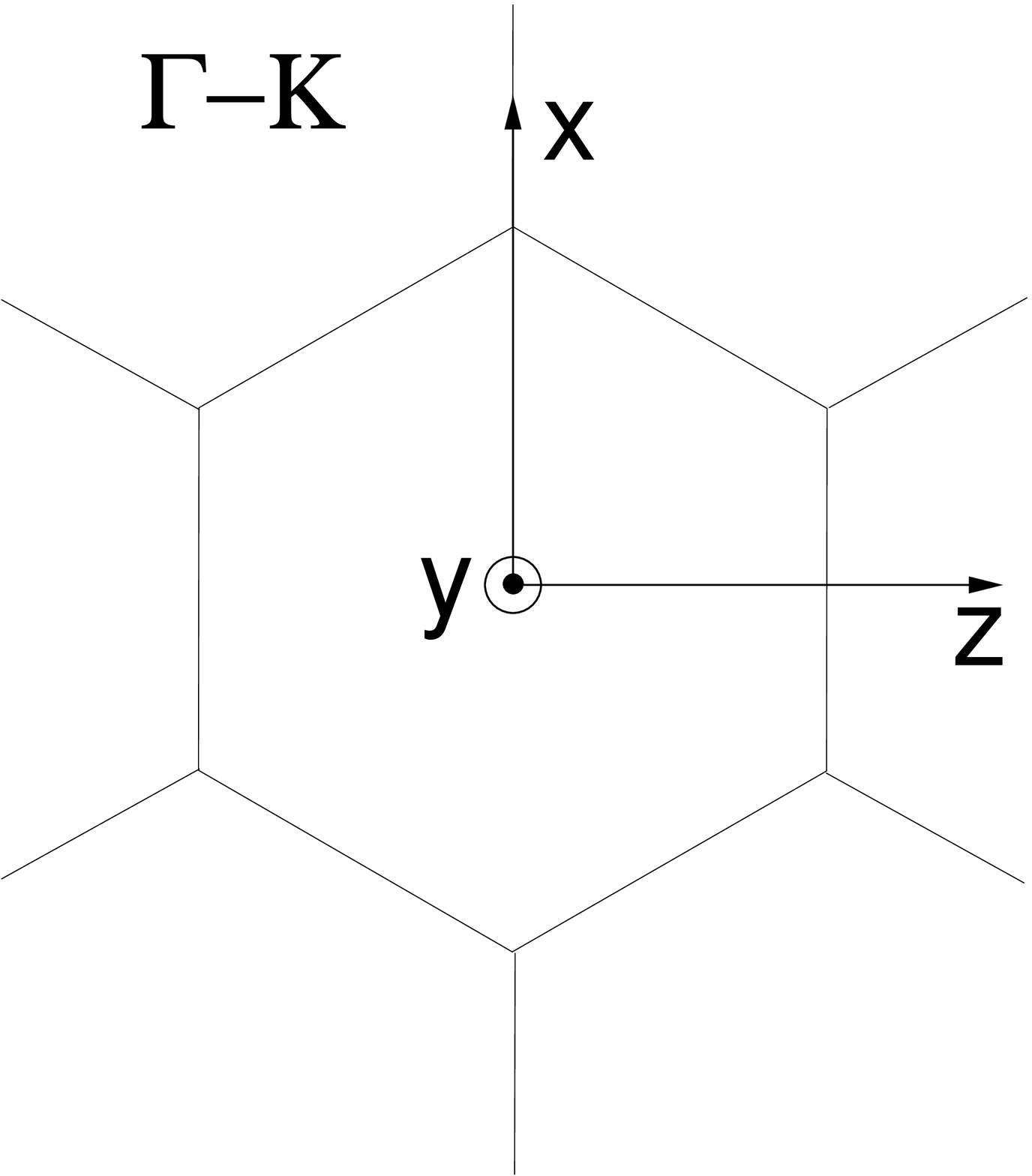,width=2.5cm}
\caption{\label{coord} Coordinate system for the point group $C_{2v}$ of $\Gamma-M$ and $\Gamma-K$ drawn in real
  space. $z$ points in the
 direction of the principal rotational axis.
}
\end{figure}

\begin{table}[htbp]
\caption{Symmetry relations between space group and molecular notation for the
  space group P$6_3/mmc$ and the point group $D_{6h}$. The corresponding coordinate system for $C_{2v}$ is shown in
  Fig.~\ref{coord}. }
\begin{ruledtabular}
\begin{tabular}{llllllll}
$\Gamma$ & $D_{6h}$ \quad $$ &        $K$ & $D_{3h}$ \quad $$ &                 $M$ & $D_{2h}$ \quad $$  &           $T,\Sigma$ & $C_{2v}$ \\ \hline
$\Gamma_2^-$  &   $A_{2u}$  &   $K_1$  &   $A^\prime_1$  &               $M_1^+$  &   $A_{1g}$  &    $T_1$, $\Sigma_1$  &   $A_{1}$ \\
$\Gamma_4^+$  &   $B_{2g}$  &   $K_2$  &   $A^\prime_2$  &               $M_2^+$  &   $B_{1g}$  &    $T_2$, $\Sigma_2$  &   $A_{2}$ \\
$\Gamma_6^-$  &   $E_{1u}$  &   $K_5$  &   $E^\prime$  &          $M_2^-$  &   $B_{1u}$  &    $T_3$, $\Sigma_3$  &   $B_{1}$ \\
$\Gamma_5^+$  &   $E_{2g}$  &   $K_6$  &   $E^{\prime\prime}$  &  $M_3^+$  &   $B_{2g}$  &    $T_4$, $\Sigma_4$  &   $B_{2}$ \\
           &           &            &            &      $M_3^-$  &   $B_{2u}$  &           &            \\
           &           &            &            &      $M_4^-$  &   $B_{3u}$  &           &            \\
\end{tabular}
\end{ruledtabular}
\label{tab:spacegrp}
\end{table}

\subsection{Fitting procedure}
\label{sec:fitting-procedure}

To obtain the optimal fit of the set of force constants to the experimental data,  we applied a variable neighbor search
(VNS) method of global optimization\cite{drazic06}, using the simplex
method~\cite{press86} for the local optimization subroutine.
Basically, this is the least squares procedure minimizing the
average deviation $\Delta(\mathbf{f})=\frac{1}{N^{\mathrm{exp}}}
\sqrt{\sum_i|\omega^{\mathrm{exp}}_i-\omega_i(\mathbf{f})|^2}$ between 
the $N^{\mathrm{exp}}=96$ experimental values, $\omega^{\mathrm{exp}}_i$,
and the corresponding calculated frequencies, $\omega_i(\mathbf{f})$, obtained
by calculation with the trial values of the force constants
$\mathbf{f}=(f_1,\dots,f_F)$. 
Due
to symmetry, for each level of the 3 or 6 neighbor atoms the same triple of force
constants can be used. 
We used stretching, out-of-plane and
in-plane force constants $f_i$ for each relevant pair of atoms.\cite{jishi93,dobardzic03} 
Here, the direction of stretching corresponds to the line that connects the center atom with the atom of appropriate
level. The in-plane and out-of-plane
directions are perpendicalur to this line and lie in the graphene layer or perpendicular to it, respectively. After
transforming the stretching and in-plane force constants to a global basis, one obtains the
dynamical matrix.
The level of the relevant neighbors
has been gradually increased until the satisfactory agreement
($\Delta<0.23$meV, with the greatest difference of
$|\omega^{\mathrm{exp}}_i-\omega_i(\mathbf{f})|\approx 8$\,meV for the LA
branch in the $\Gamma-K$ region, nearby the $K$-point) has been
eventually achieved with included neighbors of up to the fifth
level. The fifth level contains  24 neighbors of each atom, as there are 3, 6,
3, 6, and 6 symmetrically positioned first to fifth neighbors,
respectively, resulting in fifteen independent variational parameters
$f_i$. 

\begin{figure}[t]
\centering
\epsfig{file=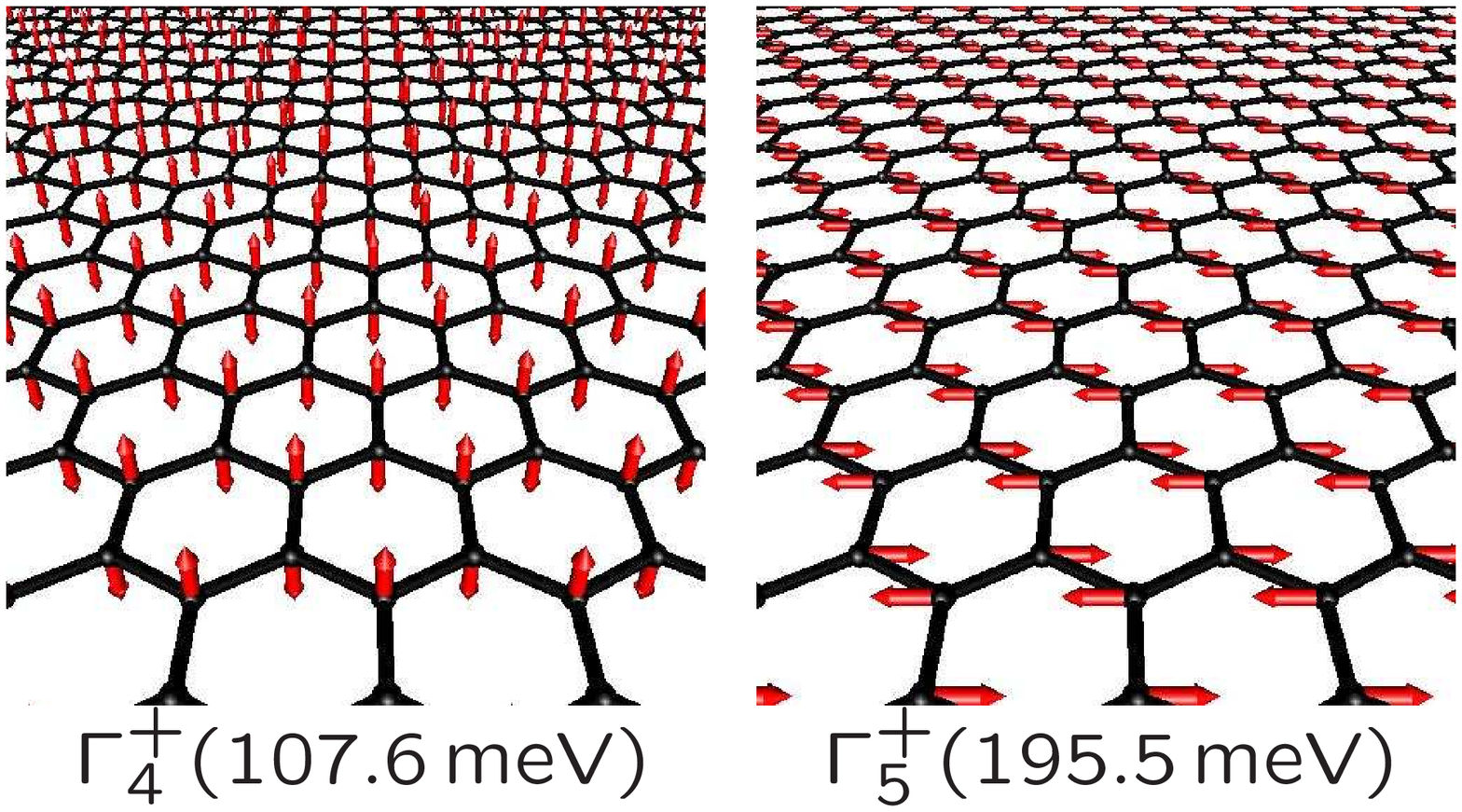,width=0.8\columnwidth}
\caption{\label{bild:Gamma} (Color online)
Optical eigenmodes of graphene at the $\Gamma$ point from the force-constants calculation. The experimental frequency values are given in brackets; they are taken from the data recorded closest to the $\Gamma$ point. 
}
\end{figure}

\begin{figure}[t]
\centering
 \epsfig{file=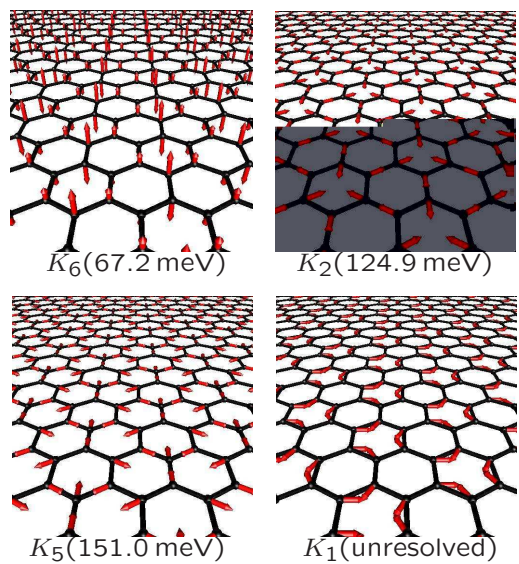,width=0.8\columnwidth}
\caption{\label{bild:K} (Color online)
Eigenmodes of graphene at the  $K$ point. For the degenerate modes, only one choice per energy is given. Phonon energies given in brackets are the IXS experimental values. For the symmetry notation see Table~\ref{tab:spacegrp}.
}
\end{figure}

\begin{figure}[t]
\centering
 \epsfig{file=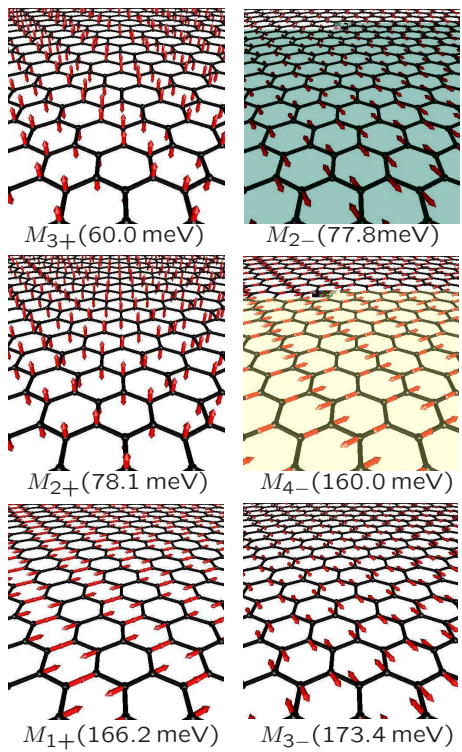,width=0.8\columnwidth}
\caption{\label{bild:m} (Color online)
Eigenmodes of graphene at the $M$ point (from lower to higher
frequencies). Phonon energies given in brackets are the IXS experimental values. For the symmetry notation see Table~\ref{tab:spacegrp}.
}
\end{figure}

\subsection{Force-constants results}
\label{sec:force-const-results}

The optimized values of the force constants parameters are presented in 
Table~\ref{tab:force_const}. In Fig.~\ref{bild:bands} we show the phonon dispersion obtained from these force constants in comparison to the experimental data. The largest deviations between the calculation and the experiment occur for the optical phonon branches near the $K$ point. This is probably due to the strong interaction of the near-$K$-point phonons with electrons near the Fermi level, which is not included in a force-constants model. Including more than 4th-nearest neighbors of atoms (i.e., fifteen independent parameters), however, gives a fairly good description of the local minimum of the TO-derived branch at the $K$ point. 

Moreover, although FC calculations including only fourth-nearest neighbors
provide a considerably good average fit to the experimental data, they lead
to permuted frequencies of the LO and LA-derived phonons at the $M$ point ($M^+_1$ and $M^-_4$), and to a  crossing of the LA and LO branches within the $K-M$ region ($K_5-M^+_1$ and
$K_5-M^-_4$). Therefore, at least fifth-nearest neighbors are required for a good empirical description of the graphite phonon dispersion. 

\begin{table}[htbp]
\caption{Force constants parameters for graphene, obtained from a fit to the experimental data, in eV/\AA $^2$. }
\begin{ruledtabular}
\begin{tabular}{cccc} 
  Neighbor level&Stretching&Out of plane&In plane\\
  \hline
  1 & 25.880&6.183&8.420\\
  2 & 4.037&-0.492&-3.044\\
  3 & -3.016&0.516&3.948\\
  4 & 0.564&-0.521&0.129 \\
  5 & 1.035&0.110&0.166 \\
\end{tabular}
\end{ruledtabular}
\label{tab:force_const}
\end{table}

The eigenvectors of all optical phonons from our force-constants calculation at the high-symmetry points $\Gamma ,K$, and $M$ are drawn in
Figs.~\ref{bild:Gamma}, \ref{bild:K}, and \ref{bild:m}, respectively. They are in agreement with calculations from a
molecular-based approach of Ref.~\onlinecite{mapelli99}. For the degenerate modes, we show only one choice per energy; the remaining eigenvectors can be obtained by the symmetry-group projectors. 

Often in literature the molecular notation for the symmetry
groups is
used. Therefore, Table~\ref{tab:spacegrp} shows the relation between the spacegroup notation of P$6_3/mmc$ and the molecular notation at
the  high-symmetry points $\Gamma, K$, and $M$, and the lines $\Gamma-K-M$ ($T$) and $\Gamma-M$ ($\Sigma$). The eigenvectors will help to choose
the sample orientation in future IXS experiments. The scattering cross section is zero, if the direction of the
atomic displacements and the momentum transfer  in the scattering process enclose an angle of 90\dg .

The above force constants parameters can in principle be used to calculate the phonon dispersion of carbon nanotubes, in
particular for chiral nanotubes with a large number of atoms in the unit cell that require large computational
  effort, when calculated with first-principle methods. The existence of a fourth acoustic mode (pure rotation of the
tube about its axis) and the finite frequency of the radial breathing mode have to be taken into account, see
Refs.~\onlinecite{dobardzic03,saito98b,maultzsch02ssc}. In addition, the different bond angles and lengths between the carbon atoms on the
cylinder surface depending on the chiral index must be included. We  expect that such an approach based on the empirical force constants of graphite
will give an overall  good description of the phonon bands of carbon nanotubes. In metallic nanotubes, however, the
coupling of the $\Gamma$-point and $K$-point phonons to the  electronic system will lead to different results for those modes, see for
instance the Kohn anomalies and the frequency drop of the LO phonon in metallic nanotubes.\cite{dubay03,piscanec04,piscanec07} 
Therefore, where the phonon dispersion is modified due to strong interactions between the phonons and the electrons, a force-constants model might only give an emipirical description of the phonon bands, but should be tested by DFT calculations which take electron-phonon coupling into account.

\section{Summary}\label{sec:summary}
 
 In summary, we presented the full in-plane phonon dispersion of graphite determined by inelastic x-ray scattering. The overall shape of the phonon bands confirms previous \emph{ab-initio} DFT calculations,  if special care is taken for the highest optical phonons near the $K$ point. We showed that by including fifth-nearest neighbors, the phonon bands can be well described within a force-constants model. Previous empirical models predicted only parts of the phonon dispersion correctly, since experimental data in the $K-M$ region had been missing. 
The new force constants will also improve the models of the phonon dispersion in carbon nanotubes.

\section{Acknowledgments}
We would like to thank M. Dra\v zi\'c for sharing his results  prior to publication. We thank A.V. Tamashausky for the rare single crystals of graphite. J.M. acknowledges support from the Alexander-von-Humboldt foundation. This work was supported in part by the ESRF.

\end{document}